# Multiferroic properties of oxygen functionalized magnetic i-MXene

Mingyu Zhao, Jun Chen, Shan-Shan Wang, Ming An,[*] Shuai Dong[†]

*School of Physics, Southeast University, Nanjing 211189, China*

**Abstract**

Two dimensional multiferroics inherit prominent physical properties from both low dimensional materials and magnetoelectric materials, and can go beyond their three dimensional counterparts for their unique structures. Here, based on density functional theory calculations, a MXene derivative, i.e., i-MXene $(Ta_{2/3}Fe_{1/3})_2CO_2$, is predicted to be a type-I multiferroic material. Originated from the reliable $5d^0$ rule, its ferroelectricity is robust, with a moderate polarization up to ~12.33 $\mu C/cm^2$ along the *a*-axis, which can be easily switched and may persist above room temperature. Its magnetic ground state is layered antiferromagnetism. Although it is a type-I multiferroic material, its Néel temperature can be significantly tuned by the paraelectric-ferroelectric transition, manifesting a kind of intrinsic magnetoelectric coupling. Such magnetoelectric effect is originated from the conventional magnetostriction, but unexpectedly magnified by the exchange frustration. Our work not only reveals a nontrivial magnetoelectric mechanism, but also provides a strategy to search for more multiferroics in the two dimensional limit.

[*]Email: amorn@seu.edu.cn
[†]Email: sdong@seu.edu.cn

## I. Introduction

Multiferroic materials with coexisting ferromagnetism and ferroelectricity in single phases, provide distinct possibilities for magnetoelectric couplings, more precisely, the feasibilities for electrically controlled magnetism and/or magnetically manipulated polarization [1-3]. For many cases of displacive ferroelectricity, the empty $d$ orbital is a prerequisite, coined as the so-called $d^0$ rule [4]. However, the partially filled $d$ shells are essential for magnetism of transition metal ions. Due to this obvious contradiction, single phase multiferroics with congenetic magnetism and ferroelectricity are not abundant.

Furthermore, when these magnetic or polar materials are made into ultra-thin films for practical applications, their ferroic properties will be seriously suppressed by depolarization fields, surface defects, and substrate strain/stress [5-6]. To overcome this challenge, two dimensional (2D) ferroelectrics, 2D magnets, and even 2D multiferroics, most of which are derived from van der Waals (vdW) materials, have come into scope and soon attracted a lot of research attentions [7-10]. These 2D ferroic materials provide great conveniences for device miniaturization [11-12]. So far, many 2D materials have been predicted to be ferroelectric or magnetic, some of which have been experimentally verified. For instance, $CrI_3$, $Ge_2Cr_2Te_6$, $MPX_3$ ($M$=Cr, Mn, Fe), $Hf_2MC_2O_2$($M$=Mn, Fe) monolayers or few-layers were verified or predicted to be magnetic [7, 11, 13-15], while $CuInP_2S_6$ few layers [12], hydroxyl-decorated graphene [16, 17], 1T-$MoS_2$ [18], phosphorene [19], $In_2Se_3$ [20], GeS [21-22], and $Sc_2CO_2$ [23], are verified or predicted to have intrinsic polarizations. Additionally, the chemically functionalized phosphorenes, some MXenes, doped $GdI_3$, $VS_2$, $VO_2I_2$, and $MoN_2$ were predicted to be multiferroics [24-28]. However, in most of these predicted multiferroics, the origins of polarization and magnetism are independent, and thus their magnetoelectric couplings are expected to be intrinsically weak.

Among these 2D ferroic materials, MXenes, i.e., layered metal carbides and nitrides, form a unique branch, which can be derived from their non-vdW parent phase $M_{n+1}AX_n$ by etching the $A$ layer [29]. However, the metal elements $M$ in MXenes are mostly nonmagnetic, like Sc or Ta, which are not easy to obtain local magnetic moment. To expand the family of MXenes and add more functions, a new sub-branch, so called i-MXenes, was designed, via partial and orderly substitution of metal elements [30-33]. By introducing magnetic metal elements into 1/3 $M$ sites, magnetism can be induced, which provides a route to pursuit 2D magnetism or even multiferroicity.

In this work, the magnetic and ferroelectric properties of an i-MXene member, i.e., $(Ta_{2/3}Fe_{1/3})_2CO_2$, will be studied via first principles density functional theory (DFT) calculations. Its parent phase $Ta_2AlC$ is nonmagnetic. To induce magnetism, 1/3 Ta ions are substituted by Fe [30]. Furthermore, to enhance its chemical stability at ambient conditions, the dangling bonds of surface metal ions are passivated by oxygen, leading to the stable valences of $Fe^{2+}$ and $Ta^{5+}$. These valences are just the preconditions of magnetism and ferroelectricity, for the partially-filled $3d$ orbitals of $Fe^{2+}$ and the empty $5d$ orbital of $Ta^{5+}$, respectively. Our DFT calculations indeed prove that $(Ta_{2/3}Fe_{1/3})_2CO_2$ is a 2D multiferroic system, and its magnetoelectric coupling can be strong, beyond the expectation for type-I multiferroics.

## II. Computational Methods

Our DFT calculations were performed based on the projector augmented wave (PAW) pseudo potentials as implemented in Vienna *ab initio* Simulation Package (VASP) [34, 35]. To properly describe the correlated electrons, the GGA+$U$ method was used and the on-site Hubbard $U_{\text{eff}}$ was imposed on Fe's $3d$ orbitals using the Dudarev approach for all calculations [36]. The plane-wave cutoff energy is fixed as 500 eV. A 9×5×1 $\Gamma$-centered $k$-point grid was used for the minimum cell [see Fig. 1(a)], and a 5×5×1 grid for the 2×1×1 supercell. A vacuum space of 17 Å was adopted to avoid the interaction between two neighboring slices.

Both the lattice constants and atomic positions were fully relaxed until the Hellmann-Feynman force on each atom was smaller than 0.01 eV/Å. The ferroelectric polarization is calculated using the Berry phase method [37]. In order to check the dynamic thermal stability and ferroelectric transition of $(Ta_{2/3}Fe_{1/3})_2CO_2$, the spin-polarized *ab initio* molecular dynamics (AIMD) was done under the canonical ensemble [38], using a 2×2×1 supercell. The AIMD simulations were carried out using a Nosé thermostat from 50 K to 700 K for 6 picoseconds at each temperature.

In addition, a classical spin Heisenberg model was used to estimate the magnetic transition temperatures. All magnetic exchanges and magnetocrystalline anisotropy were extracted from the DFT energies. Then, Monte Carlo (MC) simulation was performed on a two dimensional triangular bilayer lattice (24×24×2) with periodic boundary conditions, to mimic the ion sublattice. Larger lattices were also tested to confirm the results. Typically, the initial 2×10$^5$ MC steps are discarded for thermal equilibrium and the following 2×10$^5$ MC steps are retained for statistical averaging. In our MC simulation, the specific heat is calculated to determine the phase transition point.

## III. Results and discussion

### A. Ground state properties

The parent phase Ta$_2$AlC bulk is already commercially available, and then the Ti$_2$C monolayer can be obtained as other MXenes [39]. The Ta$_2$C monolayer is composed of three atomic layers: the middle C layer sandwiched by two Ta layers. Within each layer, the Ta and C ions form triangular geometry. Its structural symmetry belongs to space group *P-3m1* (No. 164).

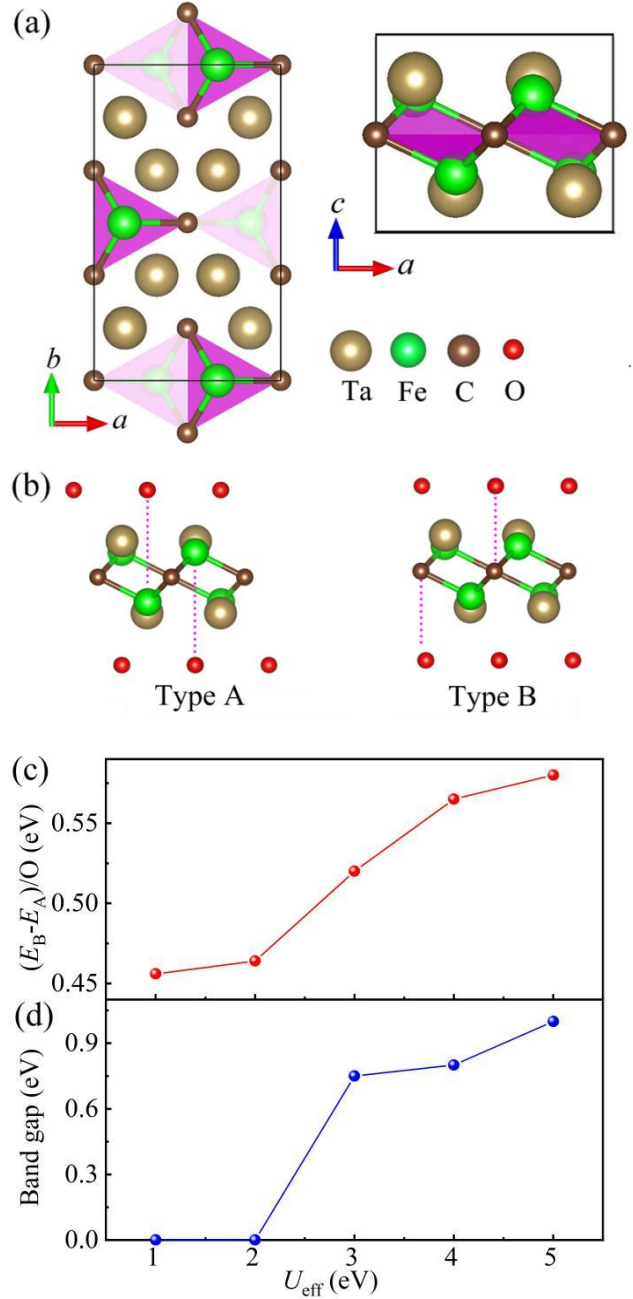

**FIG. 1.** (a) The structure of i-MAX phase of (Ta$_{2/3}$Fe$_{1/3}$)$_2$C. Both the top view (left) and side

view (right) are shown. (b) Two types of oxygen adsorption on $(Ta_{2/3}Fe_{1/3})_2C$ surface. (c) The evolution of energy difference per O between the types A and B, as a function of $U_{eff}$. The type A is always the lower energy one. (d) The band gap of $(Ta_{2/3}Fe_{1/3})_2CO_2$ (the type A oxygen adsorption and ferromagnetic state) as a function of $U_{eff}$.

As predicted in Ref. [30], the i-MAX phase of $Ta_2C$ can be obtained by replacing 1/3 Ta ions with Fe, resulting in $(Ta_{2/3}Fe_{1/3})_2C$. The partial and orderly iron substitution can not only introduce magnetism into the system, but also reduce the structural symmetry to space group $C_{2/m}$ (No. 12), as illustrated in Fig. 1(a), which breaks the in-plane three-fold rotational symmetry.

To improve the stability of $(Ta_{2/3}Fe_{1/3})_2C$, the surface metal layers are passivated by oxygen. As illustrated in Fig. 1(b), there are two most possible configurations for oxygen absorption, which lead to two structures with distinct symmetries. For the type A configuration, the adsorbed oxygen ions on one side are facing the metal ions on the opposite side. For the type B configuration, the adsorbed oxygen ions on both sides are facing the middle C ions. The energy difference between the types A and B is calculated as a function of $U_{eff}$. The result exhibited in Fig. 1(c) shows that within the considered $U_{eff}$ range, the type A adsorption is always more favorable in energy. Therefore, only the type A configuration will be considered in the following calculations. The band gaps of type A with ferromagnetism (to be further checked in the following) is also calculated, as shown in Fig. 1(d). A metal-insulator transition occurs when $U_{eff}$>2 eV.

Nominally, the 2$p$ orbitals of both $C^{4-}$ and $O^{2-}$ ions are fully occupied, while the 5$d$ orbitals of $Ta^{5+}$ are completely empty. The only partially occupied orbitals are $Fe^{2+}$'s 3$d$ ones, which lead to local moments. To determine the magnetic ground state, the ferromagnetic and four typical antiferromagnetic (AFM) configurations of Fe's bilayer are considered, as sketched in Fig. 2(a). The energy differences between these spin configurations are plotted in Fig. 2(b) as a function of $U_{eff}$. When $U_{eff}$ is larger than 2 eV, the AFM2 state has the lowest energy. According to previous theoretical studies [40-41], $U_{eff}$=4 eV is a proper choice for Fe's 3$d$ orbitals, which implies the AFM2 state. In the following, this particular $U_{eff}$ value (=4 eV) will be adopted, if not noted explicitly.

Then the electronic structure of $(Ta_{2/3}Fe_{1/3})_2CO_2$ is studied with the AFM2 order. The density of states (DOS) and element-projected DOS are shown in Fig. 2(c), and the corresponding band structures are shown in Fig. S1 of Supplemental Materials (SM) [42]. Clearly, the ground state of $(Ta_{2/3}Fe_{1/3})_2CO_2$ is a layered AFM insulator, with a band gap ~1 eV.

As mentioned before, the 5$d$ orbitals of Ta$^{5+}$ are nominally empty. According to the empirical $d^0$ rule, such empty $d$ orbitals may give rise to proper ferroelectricity via coordinate bonding, which has been well established in 3$d^0$ systems like perovskite titanates (e.g. BaTiO$_3$ and PbTiO$_3$). Recently, such $d^0$ rule was also proved to work in 5$d^0$ systems, like LiFe(WO$_4$)$_2$ and WO$_2$Cl$_2$ [43, 44]. As shown in Fig. 2(c) (and Fig. S1 in SM [42]), there is indeed strong hybridization between O's 2$p$ and Ta's 5$d$ orbitals, which is a positive evidence for the forming the coordinate bonds. As a result, there are imaginary frequencies existing at the $\Gamma$ point of its phonon spectrum for the paraelectric state, as shown in Fig. S2 in SM [42], indicating the ferroelectric instability.

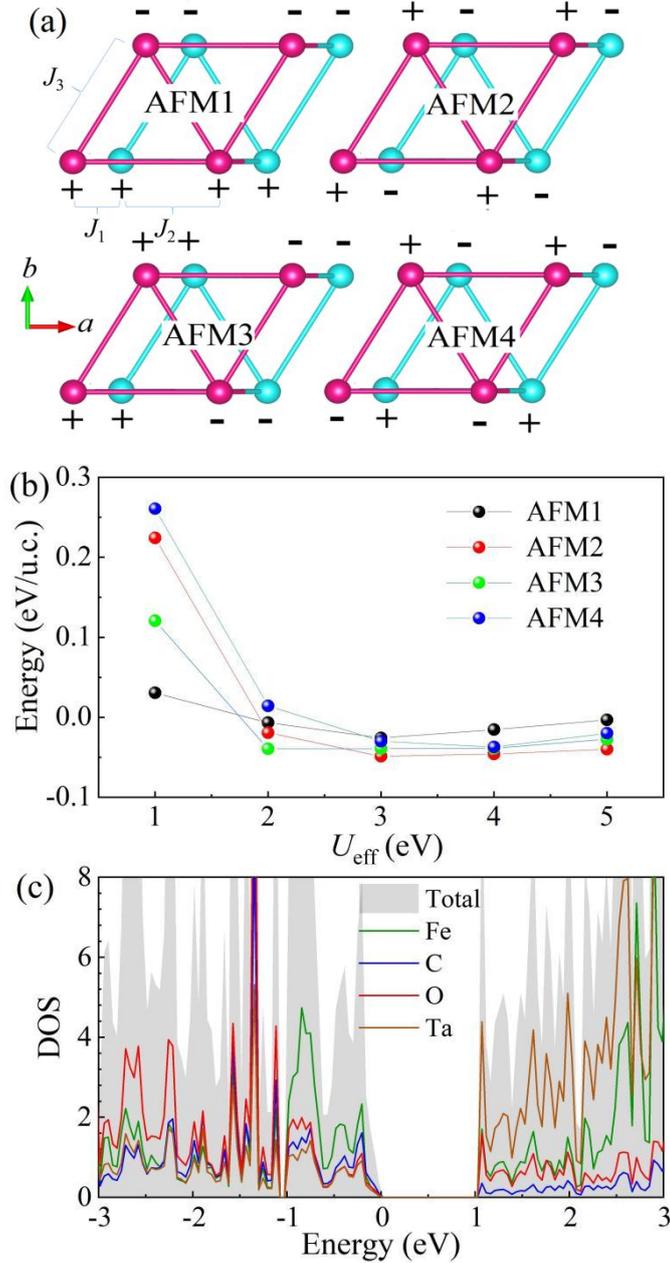

**FIG. 2.** (a) The four typical AFM configurations for bilayer triangular Fe sublattice in

$(Ta_{2/3}Fe_{1/3})_2CO_2$. The upper and lower layers are distinguished by colors, while the spin directions are denoted by signs + and -, respectively. The neighboring exchanges $J_1/J_2/J_3$ are also indicated. (b) The energy comparison of these AFM magnetic states as a function of $U_{eff}$. The energy of ferromagnetic state is taken as the reference. (c) The total DOS and element-projected PDOS of $(Ta_{2/3}Fe_{1/3})_2CO_2$, which indicate the strong $d$-$p$ hybridization around the Fermi level. In (b-c), the ferroelectric structure is adopted.

As expected, such $p$-$d$ hybridization distorts the crystal structure and further reduces the symmetry from $C_{2/m}$ to $P_a$, which belongs to the polar point group $m$. The polarization of $(Ta_{2/3}Fe_{1/3})_2CO_2$ estimated using the Berry phase method is about 12.33 $\mu C/cm^2$ (if its thickness is used to calculate the volume) along the $a$-axis. The allowed polarization can be along both the $a$-axis and $c$-axis, but can not be along the $b$-axis. However, the calculated polarization along the $c$-axis is only 0.16 $\mu C/cm^2$, which is negligible comparing with the $a$-component (12.33 $\mu C/cm^2$).

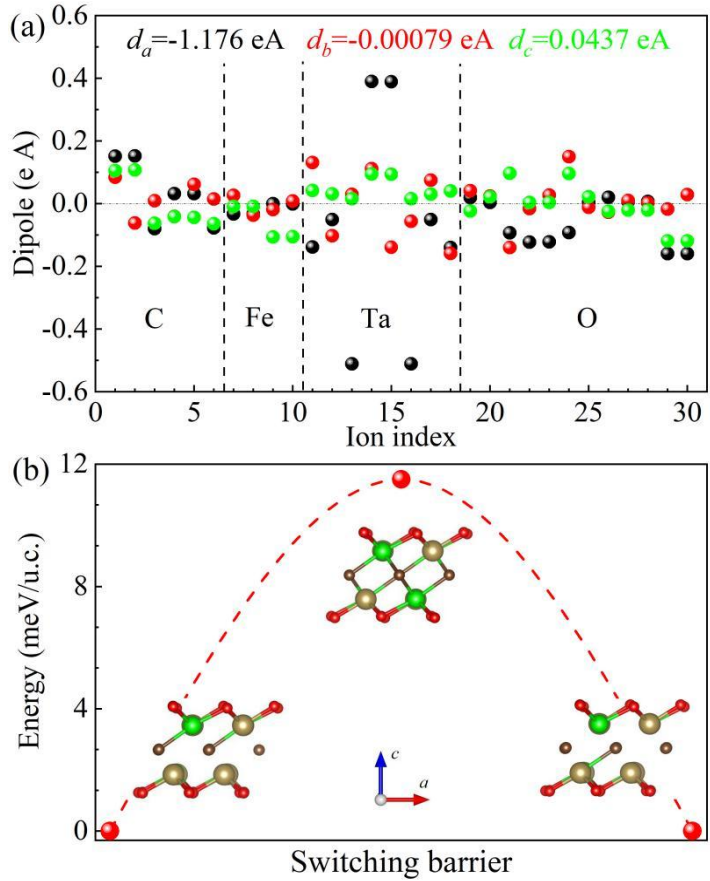

**FIG. 3.** (a) The point charge model analysis of dipole contributions to ferroelectric polarization. Each dipole is estimated using the nominal charge of ion and its displacement from the corresponding position in the paraelectric state. The $a$-/$b$-/$c$-components are distinguished by colors (black, red, green, respectively). The net dipole moments ($d_a$, $d_b$, $d_c$) of all ions indicate the dominant component along the $a$-axis. The tiny value of $d_b$, which is beyond the reliable precision, comes from the fluctuation during the structure relaxation without symmetry restriction. (b) The energy barrier of ferroelectric switching. Insert: schematic of structures.

The polarization evaluated from the point charge model is ~7.84 μC/cm² along the $a$-axis, which is qualitatively agrees with above Berry phase results. The quantitative difference between Berry phase polarization and point charge model one is acceptable and can be reasonably understood as the deviation of valences from their nominal ones, which is quite common for these partially covalent bonds [45]. In addition, the point charge model can provide an intuitional analysis of the origin of polarization, as show in Fig. 3(a). It is clear that all dipoles along the $b$-axis and most dipoles along the $c$-axis are canceled, and remanent dipoles are mostly along the $a$-axis.

The polarization of $(Ta_{2/3}Fe_{1/3})_2CO_2$ is smaller than some typical ferroelectric perovskites: $BaTiO_3$ (~26 μC/cm²) [46], $PbTiO_3$ (~80 μC/cm²) [47], and $BiFeO_3$ (~90 μC/cm²) [48], the ferroelectric polymer PVDF (~18 μC/cm²) [49, 50], but slightly larger than $Ca_3Ti_2O_7$ (~8 μC/cm²) [51], hexagonal $YMnO_3$ (~5 μC/cm²) [52], and hexagonal $LuFeO_3$ (~9 μC/cm²) [53]. In this sense, its ferroelectric polarization remains in the moderate range for potential applications.

The mechanical properties of ferroelectric films are also essential for applications. Here the stiffness tensor $C$ is calculated to verify the mechanical stability of $(Ta_{2/3}Fe_{1/3})_2CO_2$. The values of all elements of $C$ can be found in Eq. S1 in SM [42]. Our calculation results satisfy the requirements of the mechanical stability criterion of a 2D material [54]: $C_{11}>0$, $C_{66}>0$, and $C_{11}C_{12}>C_{12}^2$. The Young's modulus $Y_S$, which stands for the stiffness of material [55], is 294.4N/m for $(Ta_{2/3}Fe_{1/3})_2CO_2$. It is a little lower than that of graphene (348 N/m [56]) but higher than most of other 2D materials, such as BN (258 N/m [57]), Ge (41.4 N/m [58]), and Si (61 N/m) [58]. This means that the $(Ta_{2/3}Fe_{1/3})_2CO_2$ has excellent mechanical properties.

In the last, it is necessary to discuss briefly on the oxygen nonstoichiometry, which may occur in real process of surface passivation. The oxygen nonstoichiometry will change the valences of Ta/Fe/C. In rich oxygen environment, $Fe^{2+}$ can be further oxidized to $Fe^{3+}$, thus the ferroelectricity originated from the $d^0$ rule may tolerate more or less to these extra oxygen. In contrast, the poor oxygen environment will be detrimental to the $5d^0$-$2p^6$ hybridization between Ta and C/O, and thus suppress the ferroelectricity gradually, as occurred in $BaTiO_{3-x}$ [59].

## B. Finite temperature properties

Additionally, the AIMD simulations have been employed to verify the thermal stability of of $(Ta_{2/3}Fe_{1/3})_2CO_2$ and its ferroelectric transition. As shown in Fig. 4(a), our AIMD simulation indicates that the polarization of $(Ta_{2/3}Fe_{1/3})_2CO_2$ can persist up to 300 K in despite of thermal fluctuations. When temperature further increases to 700 K, the (averaged) polarization will be suppressed to almost zero, but its layered structure framework remains robust [also see Fig. 4(a)], implying the thermal stability of $(Ta_{2/3}Fe_{1/3})_2CO_2$. The ferroelectric transition temperature ($T_C$) is estimated as ~450 K, as shown in Fig. 4(b). Of course, this $T_C$ may be overestimated due to the finite size effect in the AIMD simulation.

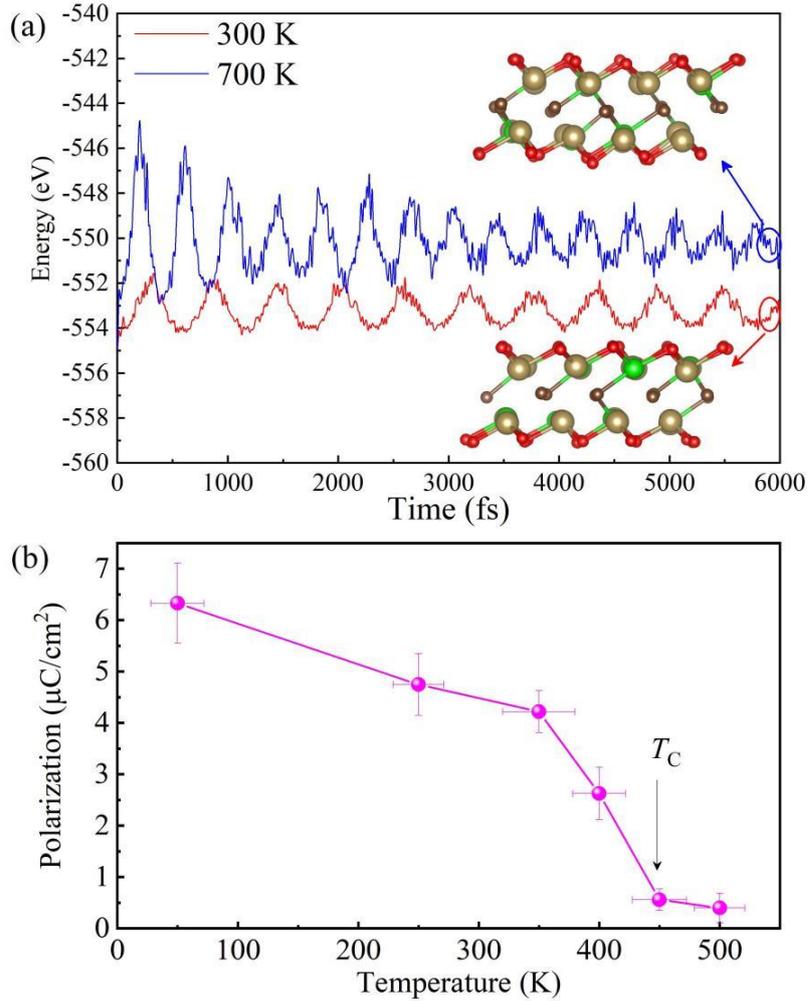

**FIG. 4.** (a) The energy fluctuations during the AIMD simulations and the final structures of $(Ta_{2/3}Fe_{1/3})_2CO_2$ at 300 K and 700 K, after 6 ps. (b) The averaged polarization estimated using point charge model as a function of AIDM temperature. The estimated ferroelectric $T_C$ is about ~450 K.

The ferroelectric $T_C$ can also be roughly estimated using the depth of ferroelectric potential well, i.e., the energy difference between paraelectric and ferroelectric states, which leads ~133 K (11.5 meV/u.c.), lower than the AIMD estimation. Of course, it should be noted that this method may underestimate $T_C$. For example, a simple benchmark of tetragonal BaTiO$_3$ leads to an even shallow ferroelectric potential well (~4.85 meV/u.c.), but its cubic to tetragonal transition in real BaTiO$_3$ occurs at ~400 K.

Besides the ferroelectric transition, the magnetic transition is another important issue for a multiferroic system. To estimate the Néel temperature ($T_N$) of (Ta$_{2/3}$Fe$_{1/3}$)$_2$CO$_2$, the MC simulation is applied to the classical Heisenberg model. The model Hamiltonian can be expressed as:

$$H = J_1 \sum_{\langle i,j \rangle} S_i \cdot S_j + J_2 \sum_{\langle i,j \rangle} S_i \cdot S_j + J_3 \sum_{\langle i,j \rangle} S_i \cdot S_j + \sum_i \left[ K_c (S_i^z)^2 + K_b (S_i^y)^2 \right] \quad (1)$$

where $S_i$ is the normalized spin ($|S|=1$) on Fe site $i$. $J_1$, $J_2$, and $J_3$ are the nearest/next-nearest/next-next-nearest neighbors exchange couplings, respectively. $J_1$ and $J_2$ are between layers, while $J_3$ is the one within each layer, as indicated in Fig. 2(a). The tiny distortion of iron triangle is neglected for $J_3$ (see Fig. S3 and Table S1 in SM [42]). $K_{b/c}$ stands for the single-ion magnetocrystalline anisotropy along the $b$-/$c$-axis, respectively.

The coefficients of $J_1$/$J_2$/$J_3$ can be extracted from DFT energies by comparing the AFM states of fixed structures (i.e., the optimized structure of AFM2 state). In particular, the energies for these states in a 2×2×1 cell (8 Fe spins) can be expressed as:

$$E_{AFM1} = E_0 + 4J_1 + 4J_2 - 8J_3, \ E_{AFM2} = E_0 - 4J_1 - 4J_2 + 24J_3,$$

$$E_{AFM3} = E_0 + 4J_1 - 4J_2 - 8J_3, \ E_{AFM4} = E_0 - 4J_1 - 4J_2 - 8J_3, \quad (2)$$

where $E_0$ is the energy base. The detail values of $J_1$/$J_2$/$J_3$ are summarized in Table 1.

The differences of $J_1$/$J_2$/$J_3$ between the ferroelectric and paraelectric states can be traced back to the distortions of Fe-C-Fe/Fe-O-Fe bonds, as compared in Fig. S4 in SM [42]. Since exchanges depend on bond lengths/angles nonlinearly and dramatically, here the changes of $J_1$/$J_2$/$J_3$ are not small, even though these bond lengths and bond angles are only slightly modified. For example, $J_2$ is the largest one since it corresponds to an (almost) 180° Fe-C-Fe bond, which is preferred for super-exchange. The ferroelectric distortion, i.e. the slight deviation (178°) from ideal 180° Fe-C-Fe bond angle, suppresses the value of $J_2$ for ~60%. In contrast, comparing with the ferroelectric phase, $J_1$ is enhanced in the paraelectric phase because its Fe-C-Fe bond angle (87.19°) is closer to 90° than that in the ferroelectric phase (86.13°).

**Table 1.** The magnetic coefficients for the ferroelectric and paraelectric state, in unit of meV.

|  | $J_1$ | $J_2$ | $J_3$ | $K_b$ | $K_c$ |
|---|---|---|---|---|---|
| paraelectric | -1.50 | 32.75 | -0.44 | -0.05 | 0.65 |
| ferroelectric | -1.13 | 12.88 | -1.59 | -1.94 | 0.20 |

The values of $K_b$ and $K_c$ can also be obtained via the DFT calculations with spin-orbit coupling, as presented in Table 1. Although both the paraelectric and ferroelectric states prefer the in-plane magnetic easy axis, the spins in the ferroelectric state are closer to the Ising-type, while they are closer to the XY-type in the paraelectric state.

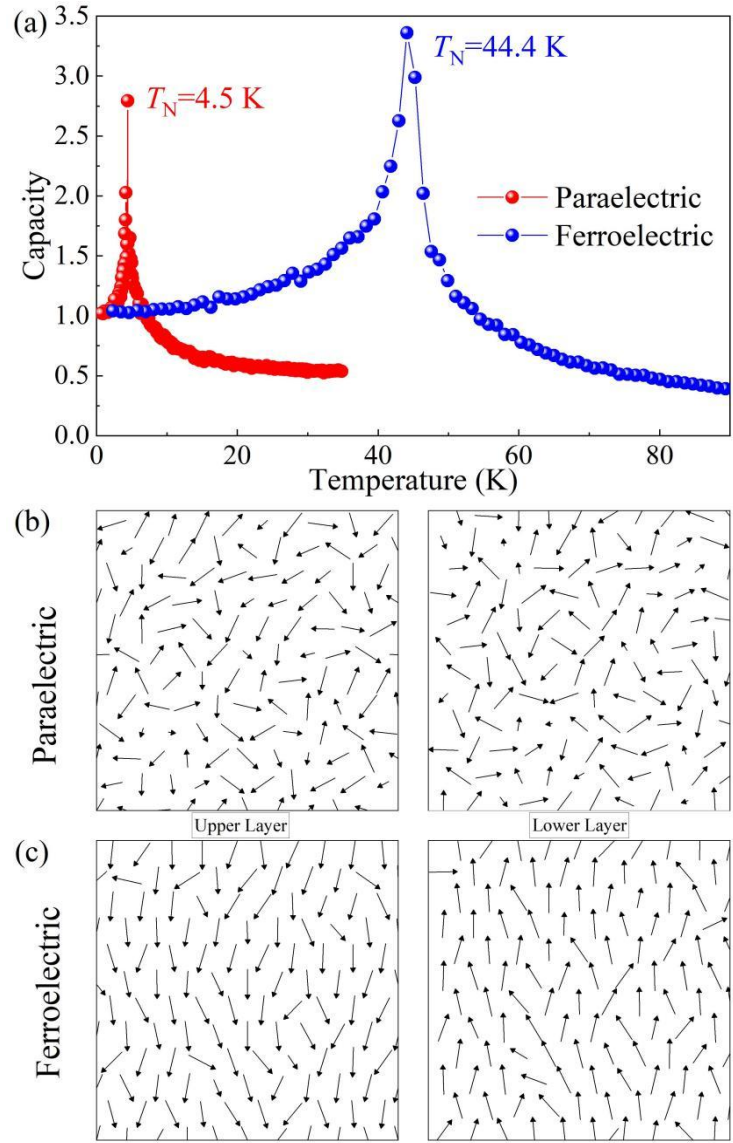

**FIG. 5.** (a) Comparison of magnetic transition temperatures (indicated by the peaks of heat capacity) of the paraelectric and ferroelectric phases. (b-c) MC snapshots of spin

configurations for the paraelectric and ferroelectric phases at 20 K. Left: upper Fe layer; Right: lower Fe layer. Only the in-plane components are shown.

Our MC simulation results are shown in Fig. 5(a). The magnetic $T_N$'s are highly different between the paraelectric and ferroelectric states: it is only 4.5 K in the paraelectric state but boosted to 44.4 K in the ferroelectric state, almost one order of magnitude higher in the latter case. This effect manifests a new type of prominent magnetoelectricity, which is unexpected for those type-I multiferroic systems.

The MC snapshots are presented in Fig. 5(b-c). Between the two $T_N$'s of paraelectric and ferroelectric states, the spin configurations are contrastive: ordered in the ferroelectric state but disordered in the paraelectric state. Although the paraelectric state is the hypothetic one at this temperature region, it can appear as the intermediate state during the ferroelectric switching, or exist as the transition state at ferroelectric domain walls. Then strong magnetoelectricity can be expected during the ferroelectric switching process and around the ferroelectric domain walls.

Finally, it is interesting to clarify the origin of such unexpected prominent magnetoelectricity. In principle, it originates from the conventional magnetostriction effect, namely the coupling between ferroelectric lattice distortions and magnetic coefficients. However, this magnetostriction effect is significantly magnified in our i-MXene case due to the exchange frustration effect, as explained below. Although $J_2$ is the strongest exchange here which is against the AFM1 and ferromagnetic states, it can not distinguish the energies of AFM2, AFM3, and AFM4 states, since all their energies depend on $J_2$ synchronously (see Eq. 2). Instead, the exchanges $J_1$ and $J_3$ will play important roles to determine the phase transition. Although $J_3$ is not large, the multiple coordination numbers of $J_3$ will make it the decisive exchange for the AFM2 state. According to Eq. 2, the energy difference between the lowest energy AFM2 and the second lowest energy AFM3 can be expressed as $8\times(J_1-4J_3)$. In the paraelectric state, the ratio between $J_1$ and $J_3$ is very close to 4:1, which leads to the effect of exchange frustration. In other words, such frustration makes the AFM2 and AFM3 very close in energy, as shown in Fig. S5 in SM [42]. As a result of magnetic frustration [60], the $T_N$ in the paraelectric state will be seriously suppressed to a very low one. In contrast, this exchange frustration is not serious in the ferroelectric state, and thus its $T_N$ can persist to a relative higher temperature.

Besides, a secondary effect is from the magneto-crystalline anisotropy. Due to the reduced symmetry in the ferroelectric state, its single-axis magnetocrystalline anisotropy $K_b$ is much stronger than that in the PE state. The Ising-type-like spins in the ferroelectric state also prefers a higher magnetic ordering temperature than the XY-type-like one in the PE state in the 2D limit.

## IV. SUMMARY

Our DFT calculations have predicted that the $(Ta_{2/3}Fe_{1/3})_2CO_2$ monolayer is a promising 2D type-I multiferroic material with unexpected prominent magnetoelectricity. Its layered antiferromagnetism is from Fe's sublattice, while the hybridization between C's/O's $2p$ orbitals and Ta's empty $5d$ orbitals leads to a proper ferroelectricity. Its moderate polarization may persist above room temperature. More importantly, a remarkable magnetoelectric effect is unexpectedly found in $(Ta_{2/3}Fe_{1/3})_2CO_2$, namely the magnetic Néel temperature can be significantly tuned between the ferroelectric to paraelectric phases, which may be realized around the ferroelectric domain boundaries or during the ferroelectric switching.

## ACKNOWLEDGMENTS

This work was supported by the National Natural Science Foundation of China (Grant No. 11834002). We thank the Tianhe-II of the National Supercomputer Center in Guangzhou (NSCC-GZ) and the Big Data Center of Southeast University for providing the facility support on calculations.

**References**

1. S. Dong, J.-M. Liu, S.-W. Cheong, and Z. F. Ren, Adv. Phys. **64**, 519 (2015).
2. M. Fiebig, T. Lottermoser, D. Meier, and M. Trassin, Nat. Rev. Mater. **1**, 16046 (2016).
3. S. Dong, H. J. Xiang, and E. Dagotto, Natl. Sci. Rev. **6**, 629 (2019).
4. Physics of Ferroelectrics: A Modern Perspective, edited by K. M. Rabe, C. H. Ahn, and F.-M. Triscone (Spinger, Berlin, 2007).
5. M. Dawber, K. M. Rabe, and J. F. Scott, Rev. Mod. Phys. **77**, 1083 (2005).
6. J. Junquera and P. Ghosez, Nature **422**, 506 (2003).
7. M. An and S. Dong, APL Mater. **8**, 110704 (2020).
8. T. Hu and E. Kan, WIREs Comput. Mol. Sci. **9**, e1409 (2019).
9. M. Wu and P. Jena, WIREs Comput. Mol. Sci. **8**, e1365 (2019).
10. X. Tang and L. Z. Kou, J. Phys. Chem. Lett. **10**, 6634 (2019).
11. C. Gong and X. Zhang, Science **363**, eaav4450 (2019).

12. S. Zhou, L. You, H. L. Zhou, Y. Pu, Z. G. Gui, and J. L. Wang, Front. Phys. **16**, 13301 (2021).
13. L. Dong, H. Kumar, B. Anasori, Y. Gogotsi, and V. B. Shenoy, J. Phys. Chem. Lett. **8**, 422 (2017).
14. B. Huang, G. Clark, D. R. Klein, D. Macneill, E. N. Moratalla, K. L. Seyler, N. Wilson, M. A. Mcguire, D. H. Cobden, and X. Di, Nat. Nanotechnol. **13**, 544 (2018).
15. K. Kurosawa, S. Saito, and Y. Yamaguchi, J. Phys. Soc. Jpn. **52**, 3919 (1983).
16. M. Wu, J. D. Burton, E. Y. Tsymbal, X. C. Zeng, and P. Jena, Phys. Rev. B **87**, 081406(**R**) (2013).
17. E. J. Kan, F. Wu, K. M. Deng, and W. H. Tang, Appl. Phys. Lett. **103**, 193103 (2013).
18. S. N. Shirodkar and U. V. Waghmare, Phys. Rev. Lett. **112**, 157601 (2014).
19. M. Wu and X. C. Zeng, Nano. Lett. **16**, 3236 (2016).
20. W. J. Ding, J. B. Zhu, Z. Wang, Y. F. Gao, D. Xiao, Y. Gu, Z. Y. Zhang, and W. G. Zhu, Nat. Commun. **8**, 14956 (2017).
21. R. X. Fei, W. Kang, and L. Yang, Phys. Rev. Lett. **117**, 097601 (2016).
22. S. X. Song, Y. Zhang, J. Guan, and S. Dong, Phys. Rev. B **103**, L140104 (2021).
23. L. Zhang, C. Tang, C. M. Zhang, and A. J. Du, Nanoscale **12**, 21291 (2020).
24. Q. Yang, W. Xiong, L. Zhu, G. Y. Gao, and M. H. Wu, J. Am. Chem. Soc. **139**, 11506 (2017).
25. L. Lei and M. Wu, ACS Nano **11**, 6382(2017).
26. J. J. Zhang, J. Guan, S. Dong, and B. I. Yakobson, J. Am. Chem. Soc. **141**, 15040 (2019).
27. N. Ding, J. Chen, S. Dong, and A. Stroppa, Phys. Rev. B **102**, 165129 (2020).
28. H. P. You, Y. Zhang, J. Chen, N. Ding, M. An, L. Miao, and S. Dong, Phys. Rev. B **103**, L161408 (2021).
29. B. Anasori, Y. Xie, M. Beidaghi, J. Lu, B. C. Hosler, L. Hultman, P. R. C. Kent, Y. Gogotsi, M. W. Barsoum, ACS Nano **9**, 9507 (2015).
30. Q. Gao and H. B. Zhang, Nanoscale **12**, 5995 (2020)
31. J. He, P. Lyu, L. Z. Sun, M. García, and P. Nachtigall, J. Mater. Chem. C **4**, 6500 (2016).
32. L. Chen, M. Dahlqvist, T. Lapauw, B. Tunca, F. Wang, J. Lu, R. Meshkian, K. Lambrinou, B. Blanpain, J. Vleugels, and J. Rosen, Inorg. Chem. **57**, 6237 (2018).
33. J. Lu, A. Thore, R. Meshkian, Q. Tao, L. Hultman, and J. Rosen, Cryst. Growth. Des. **17**, 5704 (2017).
34. G. Kresse and D. Joubert, Phys. Rev. B **59**, 1758 (1999).
35. G. Kresse and J. Furthmüller, Phys. Rev. B **54**, 11169 (1996).
36. J. D. Gale and A. L. Rohl, Mol. Simul. **29**, 291 (2003).

37. R. D. King-Smith and D. Vanderbilt, Phys. Rev. B **47**, 1651 (1993).
38. S. L. Dudarev, G. A. Botton, S. Y. Savrasov, C. J. Humphreys, and A. P. Sutton, Phys. Rev. B **57**, 1505 (1998).
39. K. Mondal and P. Ghosh, Solid. State. Commun. **299**, 113657 (2019).
40. L. F. Lin, Q. R. Xu, Y. Zhang, J.-J. Zhang, Y.-P. Liang, and S. Dong, Phys. Rev. Mater. **1**, 071401(**R**) (2017).
41. B. Z. Gao, L. F. Lin, C. Chen, L. J. Wei, J. Wang, B. Xu, C. Li, J. Bian, S. Dong, J. Du, and Q. Y. Xu, Phys. Rev. Mater. **2**, 084401 (2018).
42. See Supplemental Materials for more results.
43. M. F. Liu, L. F. Lin, Y. Zhang, S. Z. Li, Q. Z. Huang, V. O. Garlea, T. Zou, Y. L. Xie, Y. Wang, C. L. Lu, L. Yang, Z. B. Yan, X. Z. Wang, S. Dong, and J.-M. Liu, Phys. Rev. B **95**, 195134 (2017).
44. L.-F. Lin, Y. Zhang, A. Meoro, E. Dagotto, and S. Dong, Phys. Rev. Lett. **123**, 067601 (2019).
45. C. R. Gui and S. Dong, Phys. Rev. B, **102**, 180103(**R**) (2020).
46. H. H. Wieder, Phys. Rev. **99**, 1161 (1955).
47. L. Jiang, W. S. Choi, H. Jeen, S. Dong, Y. Kim, M.-G. Han, Y. Zhu, S. Kalinin, E. Dagotto, T. Egami, H. N. Lee, Nano Lett. **13**, 5837 (2013).
48. T. Choi, S. Lee, Y. J. Choi, V. Kiryukhin, S.-W. Cheong, Science **324**, 63 (2009).
49. S. M. Nakhmanson, M. B. Nardelli, and J. Bernholc, Phys. Rev. B **72**, 115210 (2005).
50. S. M. Nakhmanson, M. B. Nardelli, and J. Bernholc, Phys. Rev. Lett. **92**, 115504 (2004).
51. Y. S. Oh, X. Luo, F.-T. Huang, Y. Wang, and S.-W. Cheong, Nat. Mater. **14**, 407 (2015).
52. T. Choi, Y. Horibe, H. T. Yi, Y. J. Choi, W. D. Wu, and S.-W. Cheong, Nat. Mater. **9**, 253 (2009).
53. L. Lin, H. M. Zhang, M. F. Liu, S. D. Shen, S. Zhou, D. Li, X. Wang, Z. B. Yan, Z. D. Zhang, J. Zhao, S. Dong, and J.-M. Liu, Phys. Rev. B **93**, 075146 (2016).
54. F. Mouhat and F. X. Coudert, Phys. Rev. B **90**, 224104 (2014).
55. Q. Peng and S. De, Phys. Chem. Chem. Phys. **15**, 19427 (2013).
56. X. Wei, B. Fragneaud, C. A. Marianetti, and J. W. Kysar, Phys. Rev. B **80**, 205407 (2009).
57. A. Bosak, J. Serrano, M. Krisch, K. Watanabe, T. Taniguchi, and H. Kanda, Phys. Rev. B **73**, 041402(**R**) (2006).
58. H. Zhang and R. Wang, Physica B **406**, 4080 (2011).
59. T. Kolodiazhnyi, M. Tachibana, H. Kawaji, J. Hwang, and E. Takayama-Muromachi, Phys. Rev. Lett. **104**, 147602 (2010).
60. A. P. Ramirez, Annu. Rev. Mater. Sci. **24**, 453 (1994).